\documentclass[twocolumn]{aa}
\usepackage{graphicx}
\begin{document}
   \title{The INTEGRAL ground segment and its  science operations centre}

   \author{R. Much\inst{1},
   P. Barr\inst{1},
   L. Hansson\inst{1},
   E. Kuulkers\inst{1}, 
   P. Maldari \inst{2},
   J. Nolan \inst{1},
   T. Oosterbroek\inst{1}, 
   A. Orr\inst{1}, 
   A. N. Parmar\inst{1},
   M. Schmidt\inst{2},
   J. Sternberg\inst{1}, 
   O. R.  Williams\inst{1}, 
   \and C. Winkler\inst{1}
          }

   \offprints{R. Much}

   \institute{Research and Scientific Support Department of ESA,
   ESTEC, Keplerlaan 1, NL-2201 AZ Noordwijk, The Netherlands
   \and
   ESA-ESOC, Mission Operations Department, Robert-Bosch Str. 5, 64293 Darmstadt, Germany
}

   \date{Received ; accepted}

   \abstract{The INTEGRAL ground segment is divided into operational and
scientific components. The operational component consists of the Mission
Operations Centre, the ground stations and communications lines while
the scientific component comprises of a Science Operations Centre and 
Science Data Centre. 
The overall architecture of the ground segment is described paying 
particular attention to the tasks and functionalities of the 
INTEGRAL Science Operations Centre.
         
   }

   \authorrunning{R. Much et al.} 
   \titlerunning{The INTEGRAL science operations centre}

   \maketitle
%

\section{Introduction}

\begin{figure*}[htp]
   \begin{center}
   \includegraphics[width=12cm]{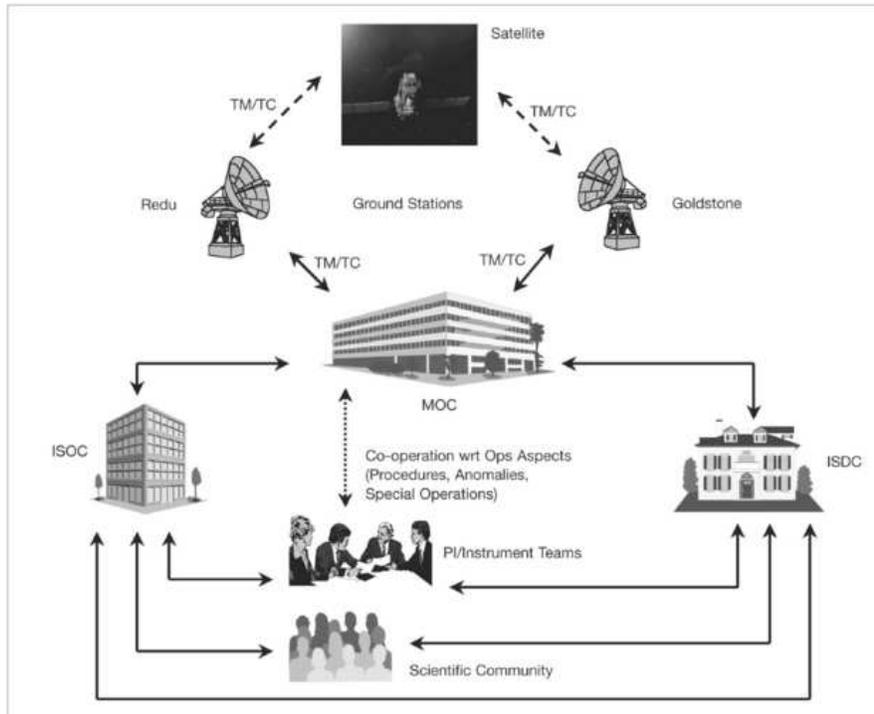}
      \caption{The overall structure of the INTEGRAL ground segment
consisting of the Operational Ground Segment 
(MOC, ground stations and communication lines) and the Science Ground
Segment (ISOC and ISDC).}
         \label{fig:structure}
   \end{center}
   \end{figure*}

The INTEGRAL Ground Segment is divided into operational and science 
components and the overall structure and interfaces
are illustrated in Fig.~\ref{fig:structure}. 
The Operational Ground Segment comprises
of the INTEGRAL Mission Operations Centre (MOC) 
located at ESOC in Darmstadt (Germany), two ground
stations and the communications system. 
The INTEGRAL Science Operations Centre (ISOC) located at ESTEC in
Noordwijk (Netherlands) and 
the INTEGRAL Science Data Centre (ISDC) located at Versoix (Switzerland)
together form the Science Ground Segment. 
The scientific community interacts directly with both components
of the Science Ground segment.
The two segments are described in more
detail below.

   \begin{figure*}[htp]
   \begin{center}
   \includegraphics[width=16cm]{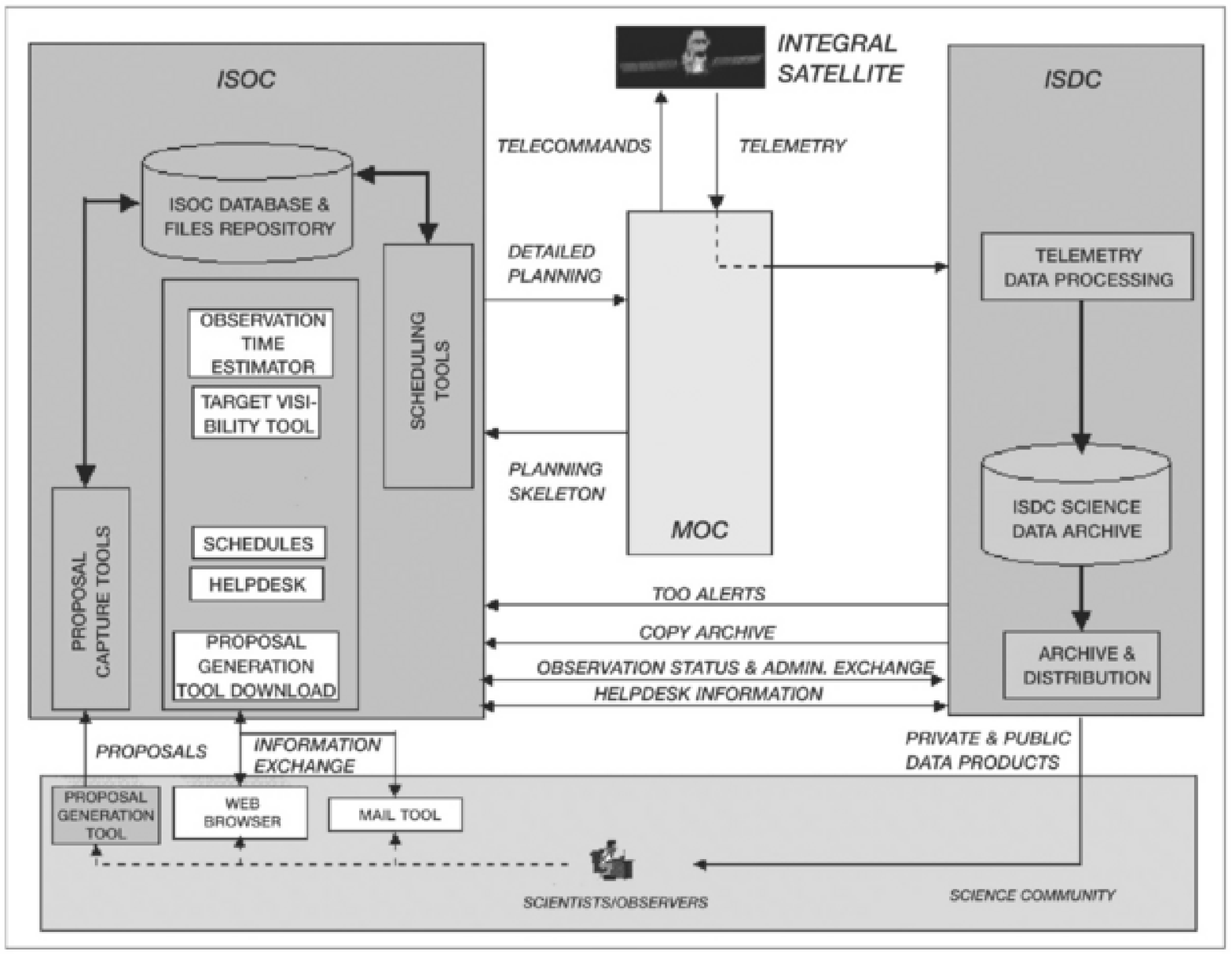}
      \caption{Data flow in the INTEGRAL Ground Segment.
The ISOC and ISDC have direct contact with the scientific community. 
The MOC is in contact with the INTEGRAL satellite via the ground 
stations in Goldstone and Redu.}
         \label{fig:dataflow}
   \end{center}
   \end{figure*}

\section{The INTEGRAL operational ground segment}

The MOC is responsible for the operation of the 
spacecraft and instruments, 
for ensuring the spacecraft safety and health, 
for the maintenance of the satellite's on-board software,
for the provision of flight dynamics support including determination
    and control of the satellite's orbit and attitude
and 
for provision of the telemetry and auxiliary data to the ISDC.
The MOC is the sole interface to the spacecraft and
all commands to be sent to the spacecraft or instruments are generated at
the MOC. 

INTEGRAL operations are executed using an automated timeline.  
The timeline is generated on a revolution by revolution basis, 
i.e. one timeline covers the period from perigee passage to 
the next perigee passage,  which is approximately three days. 
Although there are already activities much before (the planning
of ground station resources commences a year before), 
the actual  planning process starts when the MOC sends 
Planning Skeleton Files to the ISOC about one month before 
observation execution.
The Planning Skeleton Files 
define the time intervals reserved for dedicated spacecraft
operation, instrument activation and de-activation, 
handovers between the ground stations and for scientific observations.
Using these as input, the ISOC generates Planned 
Observation Sequence (POS) files
which are then sent to the MOC.
POS files contain the times when the astronomical targets selected for a
particular revolution will be observed, the target
pointings and instrument configurations, together with time windows for other
necessary spacecraft activities. The time at which an approved target
will be observed by INTEGRAL depends 
on INTEGRAL's celestial viewing constraints, 
on whether it is a fixed time observation
and 
on the overall efficiency at which it fits into the observing programme.
After a POS has been received by the MOC its correctness is verified and
the high-level information is converted into command sequences and 
parameter settings which are retrieved from an operational data base, 
which is under configuration control.
Finally, an operational timeline is generated and a summary is sent to
the ISOC for confirmation. The approved command schedule is then loaded 
into the command system at MOC during the ground station outage at the
satellite's perigee passage before the orbital revolution
concerned. The commands are then executed automatically by the command
system, and manual commanding is only envisaged under special
circumstances such as contingencies.

Since INTEGRAL is controlled in real-time from the ground, a network 
of stations has been implemented which allows permanent contact with 
the satellite during the scientifically useful part of the orbit. 
This consists of two ground stations, one located at Redu in Belgium (provided
by ESA), and one at Goldstone in California, USA (provided
by NASA/Jet Propulsion Laboratory). 
The MOC receives all satellite telemetry from the ground stations.
Based on spacecraft and instrument housekeeping data 
real-time monitoring of the operation and of the safety and health
aspects of the spacecraft and instruments are performed at the MOC. 
All telemetry and auxiliary data required to evaluate the scientific
data are routed to the ISDC in real-time.

%
The MOC maintains the onboard software and the operational database 
and is responsible for the short- and long-term archiving of the 
telemetry, telecommands and  of the auxiliary data.
The archive is regularly consolidated (filling of data gaps) 
with the satellite playback telemetry received from the ground 
stations after each pass. The consolidated data are written to CDs
within days of a particular orbit's completion and are sent to
the ISDC.

\section{The INTEGRAL science ground segment}

The science ground segment consists of the ISDC and the ISOC. Both
elements are in contact with the scientific community. 
The ISOC is mainly involved in the uplink part of the
system and the ISDC with the downlink part.

\subsection{The INTEGRAL science operations centre}

The ISOC tasks can be summarized as follows: 
\begin{itemize}
\item   to prepare the Announcements of Opportunity
for observations,   receive proposals,
assess their technical feasibility and to make these 
assessments available to an INTEGRAL Time Allocation Committee. 
\item   to assume the overall responsibility for the mission planning 
(scheduling) and for the implementation of the observing programme. 
\item to coordinate and schedule payload engineering and 
calibration observations.
\item   to assume the responsibility 
for the definition of scientific operations 
including the instrument configuration for each observation. 
\item  to  decide on the generation of an alert for Targets of 
Opportunity in order to change/interrupt the observing programme
    (responsibility of the Project Scientist). 
\item to operate the INTEGRAL help-desk in collaboration with
  the ISDC.
\item  to maintain an archive of all scientific data as created and
    populated by the ISDC.
\end{itemize}
The ISOC issues the Announcements of Opportunity for observing time. 
The proposals are submitted to ISOC using the 
Proposal Generation Tool developed at ISOC.
The received proposals are stored in the central ISOC database.
The proposals are technically evaluated by ISOC staff 
for visibility and for adequate observation duration 
using the Target Visibility Tool and the  Observation Time Estimator.
Both tools are also accessible to the science community when 
writing the proposals.
All proposals for observing time
are assessed and rated by an independent Time Allocation Committee. 
The accepted proposals are then processed at the ISOC into an optimized 
observing  plan, consisting of a sequence  of target positions
together with the corresponding instrument configurations. 
As part of this optimization process 
the ISOC checks for targets close together in the sky which can be
observed in a single observation - so saving observing time. 
This is particularly important for INTEGRAL, where the observations are
generally long and the fields of view of the $\gamma$-ray 
instruments are large. 
The optimized observing plan sequences are forwarded to
the MOC for creation of the corresponding timelines. 
Both a  long-term and a detailed short-term schedule
 are made available on the ISOC Web page (http://www.rssd.esa.int/Integral).
The ISOC routinely receives feedback from the ISDC
on the status of the executed observations. 
In case of fully (or partly) 
unsuccessful execution, rescheduling of the observation 
is considered.

The ISOC stays in close contact with the instrument teams and the MOC 
in order to coordinate any changes to the instrument configurations 
and to plan, 
coordinate, and schedule payload engineering (e.g. SPI annealing) and 
calibration  observations (e.g. OMC flat field sequences). 

A Target of Opportunity (ToO) Observation requires special treatment
by the ground segment. A ToO may be discovered using INTEGRAL itself, or
by other satellites or ground-based observatories. 
Requests for a ToO observation can be made by either 
the ISDC, or the astronomers who made the discovery,
using the dedicated INTEGRAL ToO Alert Web page. 
The Project Scientist then 
decides on the basis of scientific merit whether
to proceed with the ToO request.
If the request is granted, 
the already planned sequence of INTEGRAL observations will be 
interrupted within 20 to 36 hours and the ISOC
will generate a new observing sequence, which will be forwarded to the
MOC for validation and execution. 

There is one central help-desk dealing with all questions to the
INTEGRAL mission. The help-desk is operated jointly by ISOC and ISDC staff.
Questions on proposals, observing modes, scheduling and on INTEGRAL 
in general are handled by ISOC staff and questions on INTEGRAL data, 
instrument calibration and data shipment are handled by ISDC staff. 
This split is transparent to the user of the help desk.

The ISOC also maintains a mirror of the scientific data archive 
populated by the ISDC. 
This archive is currently restricted to internal ISOC use and it is not 
foreseen that it will be accessible by the outside world. It does, however,
also act as a backup of the ISDC archive. It would become fully
operational in case of an anomaly where
the ISDC archive is  off-line for an extended time period. 
The ISOC archive guarantees that ESA can satisfy its obligation 
to ensure the long-term availability of the INTEGRAL data.

The smooth operation of INTEGRAL is reflected in the steady progress
of the observation programme. 
Despite 
the execution of additonal PV observation in the period from 
December 17 to 30, 2002 and 
the long period spent on the essential  
Crab calibration (from February 5 to 27, 2003) 
already  44 individual AO-1 observations were observed
up to May 16, 2003, i.e. until five month after the end of the 
commissioning phase.
18 out of these 44 observations were ToO observations and 
required a deviation from the long term observation plan. 
50\% of the scheduled AO-1 time was associated with fixed
time observations. The expected number of 19 Galactic Plane Scans were 
executed and 46\% of the Galactic Center Deep Exposure time was
already spent. In total $4.8 \times 10^6${\rm s} and $2.8 \times
10^6${\rm s}
were spent on open and core programme time respectively up to 
mid May 2003.
The share of the core programme time on the total observing time
(36.8\%) is already very close to the 35\% 
it should be at the end of the first year of operation.

The overall  efficiency of the INTEGRAL operation is excellent and
higher than was expected before launch. The current scheduling efficiency is
93\%, i.e. on average 93\% of the time outside the radiation belts
is used for science observations, the remaining time is spent on slews
or reaction wheel biases. 

\subsection{The INTEGRAL science data centre}

The ISDC is responsible for searching for gamma-ray bursts and other
phenomena in the real-time telemetry, for scientific processing of the
satellite telemetry, for the archiving of the mission's scientific
products and the distribution of the data to the science community.
A detailed description of the ISDC is given by Courvoisier et 
al. (\cite{2003tc}).

\section{Summary}

The INTEGRAL ground segment is distributed over three different main
sites with well defined tasks. Operations have proceeded remarkably smoothly
since launch with 
$>$95\% of the planned operations being executed (with an average
  of 97 \%) and $>$95\%
of real-time telemetry being received  by the ISDC.
The good performance of the ground segment is contributing to
the overall scientific success of the mission and 
has validated the design of the overall system and 
has confirmed its robustness. It demonstrates 
that the INTEGRAL ground segment is well equipped to fulfill its tasks.

\begin{acknowledgements}  
The authors wish  to thank C. Breneol, N. Dean, F. Jacobs, A. Jeanes,
N. Trams, J. Treloar and R. Zondag for their support in the development and 
operations of the INTEGRAL mission at ISOC.
\end{acknowledgements}

\end{document}